# A FRAMEWORK FOR THE EVALUATION OF SAAS IMPACT

Virginia Maria Araujo, José Ayude Vázquez, Manuel Perez Cota

Informatics Department, University of Vigo, Vigo-Pontevedra, Spain

## ABSTRACT

*Nowadays the technological progress allows us to have highly flexible solutions, easily accessible with lower levels of investment, which leads to many companies adopting SaaS (Software-as-a-Service) to support their business processes. Associated with this movement and considering the advantages of SaaS, it is important to understand whether work is being developed that is underutilized because companies are not taking advantage of it, and in this case it is necessary to understand the reasons thereof. This knowledge is important even for people who do not use or do not develop/provide SaaS, since sooner or later it will be unavoidable due to current trends. In the near future, nearly all decision-makers of IT strategies will be forced to consider adopting SaaS as an IT solution for the convenience benefits associated with technology or market competition. At that time they will have to know how to evaluate impacts and decide. What are the real needs in the Portuguese market? What fears and what is being done to mitigate them? What are the implications of the adoption of SaaS? Where should we focus attention on SaaS offerings in order to create greater value? These are questions we must answer to actually be able to assess and decide. Often, decision-makers of business strategies consider only the attractive incentives of using SaaS ignoring the impacts associated with new technologies. The need for tools and processes to assess these impacts before adopting a SaaS solution is crucial to ensure the sustainability of the information system, reduce uncertainty and facilitate decision making. This article presents a framework for evaluating impacts of SaaS called SIE (SaaS Impact Evaluation) which in addition to guidance for the present research, aims to provide guidelines for the collection, data analysis, impact assessment and decision making about including SaaS on the organizations strategic plans.*

## 1. INTRODUCTION

The technological changes that have been observed, allow anyone to have access to IT infrastructures with greater processing capabilities and storage, communication networks increasingly faster and greater bandwidth from any geographical location, through several different mobile and non-mobile devices. Because of this evolution, the SaaS solutions are presented as a very interesting and solid solution, mostly satisfying the needs of its users. The question that arises now is how to identify opportunities, challenges and solutions that may emerge in response to the progress of the SaaS model. It is necessary to understand how companies can adopt this model as a new form of production, use and consumption of software, linking it with business models that they already have. How is it possible to offer software to several companies and customers simultaneously? This is the concept of long tail [1]. Furthermore, it is necessary to know how to identify all relevant aspects associated to this type of software, which may impact the organizations and thus can identify implications of using it and evaluate for a correct decision in the sense obtaining the highest value.

Vidyanand Choudhary proves the theory that the SaaS model can promote increased product development, improve the quality of software products in most cases and that IT manufacturers and producers may obtain further gains compared to traditional models of software licensing under certain conditions [12]. In addition to the optimization of enterprise data center technologies, there is now a reasonable offer of SaaS solutions. Major market players such as Microsoft, IBM, HP, Dell, Salesforce.com and Amazon are developing new SaaS solutions, adding capabilities to the offer they already have to strengthen their market positions [31]. There may be benefits to using SaaS enabling companies to focus on business aspects rather than technological challenges [32]. Gartner predicts that by 2016, at least 30 % of ERP and CRM applications are delivered through SaaS models , however , by 2014 , 60 % of customers of SaaS are exposed to interruptions due to a lack of appropriate SLAs (Service Level Agreements) [18]. Often, decision-makers of business strategies consider only the attractive incentives of using SaaS ignoring the impacts associated with new technologies. The need for tools and processes for assessing these impacts before adopting a SaaS solution is crucial to ensure the sustainability of the information system and reduce uncertainty. Existing research addresses specific aspects and few studies give a broad overview of the implications of SaaS for anyone who develops and provides software and also for those who consume it as an end user. There are studies that evaluate user satisfaction, usage and intent to use SaaS [22], models for risk assessment of SaaS [3], [11], [10], others that address the technical aspects of developing SaaS [23]. A recent study by Luoma goes beyond the technical aspects explaining them in the context of the business model [37]. Another study examines the benefits and risks of SaaS. On the several studies that were analyzed, very oriented to the users, there is a common suggestion for additional studies targeted to SaaS providers and their networks of partners [50]. Other studies have analyzed the risks on SaaS adoption, concluding that businesses companies should be driven by SaaS providers by mitigating technical and economic risks associated with a relationship based on SaaS [4]. Xin also makes an analysis of the risks of adopting the SaaS from the client side [52]. It was not found in the literature nor in the industry a study that provides information about the impacts of the SaaS, in a comprehensive manner including the perspective of the provider and also the perspective of the service consumer. A study that gives us a holistic view considering the implications in the development, provision and use of SaaS in different types of businesses (small , medium and large ) is more than justified and even more in the Portuguese context where research on this theme is still very meager. The major contributions of this study are:

- Knowledge of the implications that SaaS can have in enhancing the competitiveness and development of Portuguese companies to carefully plan relevant investments for its human capital, processes and technologies.
- The framework for assessing the impact of SaaS as part of the literature review, which in addition to guiding the present research aims to provide guidelines for data collection, data analysis, impact assessment and decision making about including SaaS on the organizations strategic plans.

## 2. LITERATURE REVIEW

At the beginning of the current millennium the term SaaS is officially introduced, completely decentralizing the use of information systems leading to a paradigm shift in the software industry, where from any part of the world and from any device you can access the information without requiring the user to install anything [5]. Until today there is no consensus about the definition of SaaS [7], [26], [20], [19], [12], [21], [45], but generally we can say that it is a form of supplying software in which the customer has rights over their data, and the use of software without needing to purchase a license or buy the software as if it were a product, and where the sharing of software by multiple tenants can take many forms with papers and responsibilities interacting in a common application environment [41]. Its intrinsic value is to smooth the investment costs for of Software maintenance [39]. Taking advantage of software and services, it is possible to maximize the choice, flexibility and capabilities of users in general [21]. However, before moving to a SaaS model organizations must obtain answers to some questions about the objectives and benefits to be achieved, if there is any need for customization and integration with local applications, security requirements and service continuity [21]. Despite the advantages of SaaS, its adoption still causes doubts in many CIOs (Chief Information Officer) complicating the decision, which is one of the obstacles to SaaS growth and expansion. Among the concerns pointed one is the service availability. Donna Scott in a joint work with Robert and Alexa Bona, call the attention to the costs of unplanned SaaS downtimes [51]. The contractual guarantees and SLAs agreed with the SaaS provider, to mitigate the risk of service disruption are important. The required level of service uptime (availability) depends on the criticality of the application for the organization [51]. The option of integration is important to expand the SaaS offering among large companies. Increasingly integration is necessary between internal and external business systems for each organization. To meet the business needs, sometimes it is necessary to integrate SaaS applications with on-premise applications locally installed [35]. The integration and customization are critical components in the strategies of successful SaaS architectures, a centralized IT infrastructure based on services [9]. In many cases, this means creating dependencies of synchronization and data transfer between the SaaS solution and one or more internal applications [8]. SaaS customers should set their own strategies for integrating SaaS and should incorporate them in a holistic approach to multi-enterprise integration [33]. The features and quality that each particular client requires a software solution can be naturally different. As a result, SaaS providers need to answer the specific needs of a broad spectrum of potential tenants enabling the configuration and customization for each specific tenant [40]. For SaaS developers and providers, the integration and customization are other challenges to face since the opening of ports in the firewall and dedicated VPNs approaches have limitations and drawbacks, which result in many companies don't feeling comfortable in running business-critical applications on the other side of the corporate firewall [36], and using SaaS to support critical processes within the organization [16]. Regarding SaaS architecture, there are several doubts and questions about the most effective model to use. In 1995, Gacek and Boehm Allah, make a study of the various definitions of software architecture that confirms that software architecture comprises set statements about the system user needs, an architectural logic that demonstrates that the components, connections and constraints define the system when implemented, meet the user's requirements [25]. Desisto and Paquet present four architectural models used by most SaaS developers and providers [19]. The application must be customizable, configurable and must have multi-tenant efficiency [7]. The existence of a transport service layer SOAP, Web Service with XML and relational network security protocols are crucial to ensure the accuracy of the analysis and delivery of information [34]. With respect to scalability, "a measure of the ability of an application system to, without modification, provide higher response time considering the cost-benefit and/or support more users when more hardware resources are" is a complex problem in distributed systems and is another challenge for SaaS providers [54]. Data security is crucial for any information system not only for SaaS solutions. This is a major concern for SaaS users and one of the largest challenges for software architects. Even those who use or are considering using cloud services, put safety at the top of their list of concerns, ahead of performance and application availability [15]. To research this theme the new strategy of the European Commission: " Unleashing the potential of cloud computing in Europe " outlines actions to achieve a net gain of 2.5 million new jobs in Europe, and an annual increase of EUR 160 billion to EU GDP in 2020 [24]. In this type of applications where multiple users share the same resources, the best way to protect data, is through logical partitioning of data and configuration. Thus, based on the identification of the client security is guaranteed multi-tenancy [43]. The degree of isolation of a SaaS application may vary significantly depending on the business requirements and there are different techniques for implementing flexible schemas for SaaS. There are opinions that it should be possible to dynamically evolve the schema of the database, online and in "self-service" mode for each tenant, otherwise the SaaS provider has very high operational costs [2]. Authors like Aulbach, Jacobs, Kemper, & Seibold suggest techniques such as the use of XML columns and Pivot Tables [2]. Major players in the SaaS market are developing variations on the multi-tenancy architecture. Example is the new database from Oracle 12c available as an option that allows users to lodge several databases "pluggable" within a single host-based. Oracle says that this approach is more secure than multi -tenancy at the application level used by other players such as Salesforce.com. Providers like Oracle and SAP are now perfecting their knowledge into something that other SaaS providers such as Salesforce.com and NetSuite had to do earlier: How to manage a software company based on annual subscriptions and not licenses of perpetual software? [30]. Confirming the prediction that PaaS will become a "must-have" software for business class and there can be no SaaS PaaS [15], the Salesforce.com announced last November the launch of Salesforce1 platform, improved its initial offer Force.com with the tools of the purchase of Heroku and ExactTarget with mobile support and 10 times more APIs than the previous version. This is a challenge for competitors and for academic researchers to make available open standards, providing knowledge to the general public so that other companies can compete and establish itself in the market that is not only for those that have great financial ability to investigate and explore new solutions, and thus counteracting the tendency of large providers undermine and absorb small providers in economies of scale [16]. Another aspect to consider about the development of the SaaS is the business model to use and that it may simply involve higher costs and greater efforts to build the software, representing a new challenge for the design of payment models [36]. The change in business model may involve several aspects [38] including those related to human resources. Much remains to be done on these topics that are factors to consider in the design of any software application and also SaaS applications. Some studies conclude that SaaS is an immature model of Software delivery [4]. To deliver software as a service rather than as a software product requires software providers and telecom operators to change the way of thinking in three interrelated areas: business model, software architecture and operational structure [2]. The role of the providers has to radically change from a remote application storage, to an active agent in the management of a complex software ecosystem, where all the IT resources required are coordinated in order to maintain and create value for all the parties involved [50].

## 3. METHODOLOGY

One of the objectives of this study is the involvement of managers and decision makers from companies in the Portuguese market, to assess the impacts of the adoption of SaaS solutions in their organizations, and confirm the hypothesis that SaaS brings benefits and enhances their development. Place a gathering feedback both from those who use and consume SaaS but also from those who develop and/or provide SaaS, thus covering the two points of view in the software trade, to realize what actually constitutes a problem in SaaS and that effectively companies would like to have available in this offer and what their implications are. To this end, it is concluded to be more appropriate to adopt a research strategy based on a combination of methods. By combining methods, complementing a disadvantage of one method with the advantages of others, increases the reliability and significance of the research. A combination of a quantitative method with a qualitative method approach complements and enriches the research [28]. The researchers feel that the biases inherent in one method may counteract the prejudices of other methods. From the original concept of triangulation of different data sources, crosses are coming in quantitative methods with qualitative methods [13]. This study uses a research strategy based on a combination of methods sequentially. First a survey is used to get a broader view and to provide baseline information through a quantitative research, allowing subsequently to have a greater depth in quantitative research, helping to interpret and contextualize the qualitative results. The results of this first quantitative research complement the qualitative research that seeks to understand the different types of Portuguese companies according to the perspective of the decision-maker, manager or director within its particular operational context.

## 4. SaaS Impact Evaluation Framework

In the conducted literature review, there was no reference to frameworks, standards, guides or even specific processes for assessing the impact of SaaS in business, simultaneously involving the perspectives of the SaaS providers and SaaS consumers. Thus, reflecting the perspectives of this literature and in accordance with their recommendations, we propose a framework that addresses the different perspectives of the purchaser-provider service, evaluating specific categories of the impact of SaaS. We call this framework SIE (SaaS Impact Evaluation) and for its preparation we based us on the fundamental concepts of ITIL (Information Technology Infrastructure Library) [48] and PMBOK [46]. We adopt the ITIL framework as a reference for the following reasons: First, it is a comprehensive framework of processes, technology, people, and understands the logic of the impact assessment of new services in the organization's service portfolios [29]. Second, it is recognized both by professionals and by scholars [6]. Currently, ITIL is a standard for information technology management that enables the integration of business with IT through the application of a method process-oriented service. It's now used by many hundreds of organizations around the world and provides the best practice guidance applicable to all types of organizations to provide quality on IT services, processes, functions and other necessary resources to support them (Official ITIL Website , 2014 ). The framework that we propose, in addition to guiding this research, aims to facilitate evaluation of the impact of SaaS in organizations, based on this, review if the specification of a new service corresponds or not to the business needs. Based on this assessment, decision makers of the business and technology strategies can make informed decisions that may be acceptance and advancement for implementation in accordance with the defined specification, or to review and reassess:

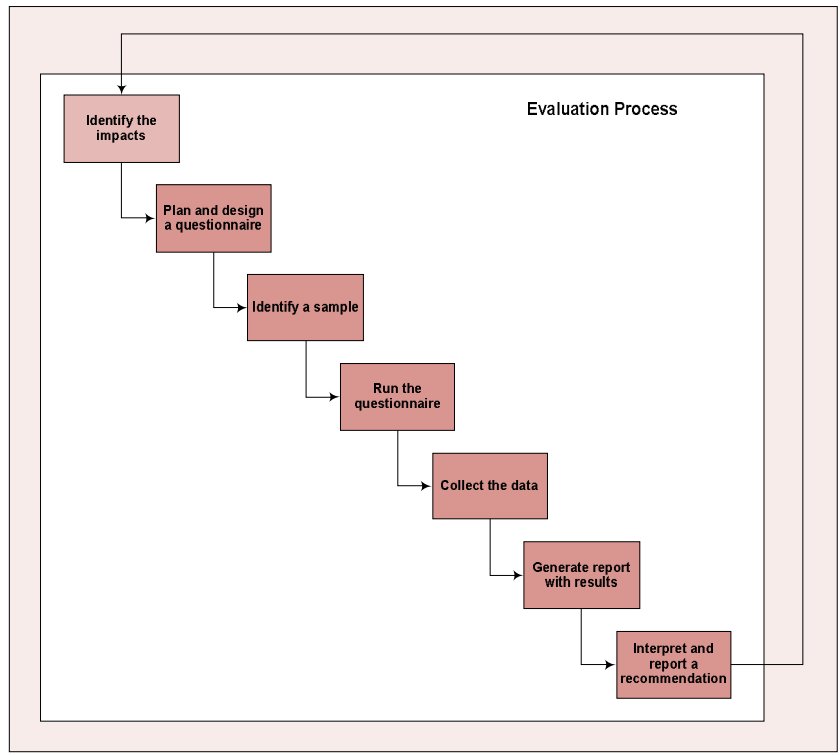

Fig. 28 SIE (SaaS Impact Evaluation)

To assess the impact of SaaS, first it is important to know the stakeholder´s goals and needs. This corresponds to the first step of a **7 step process**:

    Step 1: Identify the impacts
    Step 2: Plan and design the questionnaire
    Step 3: Identify the sample
    Step 4: Completing the questionnaire
    Step 5: Collect the information
    Step 6: Generate the report with results
    Step 7: Interpret and report with the recommendation of tasks to implement

- **Step 1: Identify the Impacts**

Identify the relevant aspects of a specific SaaS solution, and then assess their impacts on Portuguese organizations for making the final decision. Applying this to the present study, we describe the method and the data identification.

**Method:** To identify the categories and descriptions of SaaS with the most relevant impact on the business models of the Portuguese organizations, first an initial descriptive framework is prepared, identifying the non-functional and technical aspects, based on the literature. Secondly, a quantitative method to develop a broad and comprehensive study of issues to be extended is used for the collection of data from several Portuguese companies. To understand the different types of Portuguese companies from the perspective of the decision maker, manager or director within its particular operational context, in its reality, the study is complemented with a qualitative approach, thereby increasing the relevance of research and enriching the framework with more specific and actual information confronted with the theoretical assumptions of the study base.

**Identify the Data:** The main considerations derived from the literature review and also from the industry are the basis for drawing up the initial framework showing the impact of SaaS considering the research objectives.

| Category | Description |
| --- | --- |
| Company Profile | Characterization of the company type. If it is an SME, large enterprise, technology-based or not. Knowledge level of the respondent, on the concept of providing software as a service. |
| Market Positioning | Importance of SaaS offer to ensure the positioning of companies in the market, level of satisfaction with the solutions and the main difficulties in the enterprise that limits the use of this model. |
| Revenue and Profit | Gains obtained through the use of SaaS solutions and perspectives for the future. |
| Levels of Functionality, Updates and Roadmap | Configuration options, the degree of adaptation to the needs of each company, perspective of evolution and service continuity. |
| Free versions and Training | Information on types of versions and level of acceptance by end users. |
| Subscription models and SLAs | Existence of different subscription options. Lack of options due to the uniformity of the offer or evaluate if this is a difficulty in the design and development of SaaS. |
| Hybrids models | Existence of online and offline hybrid models in SaaS solutions. |
| Architecture and R&D (Research and Development) | Categories of concern in the development or use of SaaS. Interoperability and integration with other applications in the organization, architecture Multi-Tenancy, etc. |
| Infrastructure models | Architecture models, In-house, External, PaaS, IaaS, etc. |
| Security | Security level, confidentiality and privacy guarantees of user's data, security mechanisms and tools. |

- **Step 2: Plan and Design the questionnaire**

At this step, in the present study we describe the research method and also the research design.
**Method**: The big question of this research is: *What is the impact of SaaS on the development and growth of Portuguese companies?*

From this big question arise the following research questions:
RQ 1: What are the real needs in the Portuguese market?
RQ 2: What fears and what is being done to mitigate them?
RQ 3: What are the implications of the SaaS adoption?
RQ 4: Where should we focus our attention on SaaS offering in order to create a greater value?
These issues delimit the object of this study: *The impact of SaaS on models and business strategies of enterprises, to contribute to their economic recovery through the development and growth of business in a specific region of depressed economy.*

Making a brief review of research methodologies, it was concluded that to answer the first two questions a quantitative method would answer. However, the third and fourth questions require an analysis more contextual and closer to the operating reality of the companies. Thus, we arrive at the conclusion that a combined method of a quantitative and qualitative analysis is the most appropriate [13].

**Research design**: The study starts with a quantitative analysis of data from two online surveys sent to a sample of Portuguese companies to evaluate the SaaS adoption level, identifying gaps and opportunities that require further analysis in the enterprise Portuguese market. One of the surveys is sent to different types of businesses (small, medium and large) based in Portugal, which are developers and/or providers of SaaS. The second survey is sent to companies that adopt one or more SaaS solutions in their organization, in a perspective of users. The results of this first quantitative research complement the qualitative research conducted, which aims to understand the different types of Portuguese companies according to the perspective of the decision maker, manager or director within its particular operational context. In this second case, data is collected through semi-structured interviews face-to-face with directors/managers responsible for the organization IT strategy or business area [55]. Based on the results of the quantitative research, achieving the answers to research questions RQ1 and RQ2, are elaborated two interview guides with new, more specific questions to the object of the study. One of the interview guidelines is directed to SaaS developers and/or providers and another directed to SaaS users. Each interview is audio - recorded and lasted about an hour. During the qualitative interviews, notes and comments are taken by the interviewer/researcher. All the interviews are processed identically to allow identification of the existence or not of the standard deviations and what factors can explain these deviations.

## Step 3: Identify the sample

The enquired sample should involve several people from different areas of the organization to ensure that the information is reliable and consistent. Therefore and because we intend to evaluate the impact of SaaS considering the different perspectives of providers and service consumers, we selected Portuguese companies of different sizes, SaaS developers/providers and also SaaS business users or potential users.

## 2.4.4 Step 4: Completing the questionnaire

For quantitative analysis an online survey is used to get answers for the first two research questions (RQ1 and RQ2). Next face-to-face interviews are conducted to complement the answers to the last two questions (RQ3 RQ4). At this step, in addition to the aspects identified in the quantitative analysis, we consider other aspects related to the context, perceptions obtained from the contact with the real environment of the analyzed company.

### 2.4.5 Step 5: Collect the information

At this step we proceed with the collection of the information that should be stored in a safe, central and accessible location for further analysis by the various stakeholders involved in the interpretation.

Following the recommendations of [55], before starting the narrative we should identify the narrative topics. Thus, issues of online surveys and semi-structured interviews are numbered, grouped and organized according to the descriptive framework, to allow the analysis of the cases. The transcript of each interview is sent to the participants to confirm the information provided during the interview, to eliminate any misunderstanding. According to [44], we should reduce the volume of raw information and identify significant patterns, to build a communication structure of the essence revealed by the data. Quantitative data are analyzed using statistical tools that enable its analysis. The analysis of qualitative data, though it begins during the execution of the interviews starts in fact at the encoding of the collected information phase. The best way to assess results is to identify the different patterns in the collected data. For the analysis of qualitative data we can use software for qualitative data analysis.

- **Step 6: Generate the report with results**

We proceed with the creation of the report that shows the results of the questions answered from the various players, making clear the impartiality of the evaluator.

Qualitative content analysis reveals patterns, themes and categories whose end result is a descriptive framework for SaaS based on the interpretation of the researcher of those collected and coded data [44]. At this step comparisons are made between the cases and the intersection of information, to identify what is common to all cases and what is specific to a particular case. Through this data analysis, the patterns are identified and their deviation. The results of this step are the objective findings from the analysis of the collected data.

- **Step 7: Interpret and report with the recommendation of tasks to implement**

At this step we proceed to the interpretation of the results, identifying the positive and negative impacts of the presented recommendation.

In addition to the descriptive framework with aspects of greatest impact on the organization that is doing the assessment identified on step 6, the interpretation of the researcher on the results of the analysis is presented [44]. The interpretations of the researcher cannot be separated from their own training, history, context and previous understanding [14]. This step closes the process of evaluating the impact of SaaS with identifying their implications in the organization and issue a final recommendation. This recommendation facilitates the decision process to proceed with the implementation of the service considering the requirements identified or instead of the implementation as it is specified, reshape the requirements and restart the evaluation process. On the basis of this evaluation process is the Plan-Do-Check-Act (PDCA) model to ensure consistency between reviews. Such as ITIL and other Standards are based on this model, we also use it on the SaaS impact evaluation framework that we propose [17].

**Decision Process**

On the basis of substantiated information resulting from the evaluation process, the decision to proceed or not with the implementation or adoption of SaaS within the analyzed specification parameters, will always seek to maximize the value of the selected software, which have a minor impact either technological or organizational. A good and assertive decision process is based on a wide variety of scenarios to analyze and determine the most appropriate solution from the available options and reduces making wrong decisions. In choice tasks involving a large number of options, the pattern of information of a decision maker can be based on attributes to reduce the set of choices, or standards-based options to make a final decision [49] where the individual and context preferences affect the way a situation is perceived [27]. For decision-making we should reduce the uncertainty that depends on the information that we have. The best-known approach is the qualitative or quantitative assessment of factors and computer-based science is sought to provide models in order to help decision-makers [42]. The assessment framework proposed in this paper, considers that the use of a decision process should be normative in the context of SaaS evaluating, meaning that the decision process defines how the decision maker must execute the decision process. However, no reference is made as to the decision-making model to adopt. It is recommended using any decision model to help decision makers to be more rational.

## 5. SIE RESULTS VALIDATION AND DISCUSSION

The validation of the proposed framework was initially applied to the present study for the different SaaS solutions of each companies included in the research sample, allowing us to achieve the answers to the research questions. The four research questions were adequately answered based on the literature review, based on the analysis of collected data. We have the confirmation of the SaaS descriptive framework as a reliable tool for the SaaS impact assessment process, through the convergence of perspectives from who provides and also from who consumes SaaS. The identification of the major concerns in different dimensions - technical, business, organizational / operational answered the first two questions. The literature has revealed several risk factors associated with adopting SaaS. The data collected shows that some of these risk factors are associated with fear and mistrust in the model. However, the applicant's security concerns, integration, customization and business continuity were the predominant responses from those who want to consume as part of who develops and distributes SaaS. This research adopted several of these concerns as factors worth assessing the impact of SaaS to prove its legitimacy with a quantitative study. The third research question on the implications of SaaS adoption was answered with an analysis and interpretation of collected data. The generic answer to the third question is that the decision makers of the business strategies of Portuguese companies are aware of the technological and organizational impacts associated with SaaS adoption. The fourth and final research question sought a solution that connects mitigating the impact of SaaS with the decision process. The process of impact assessment and the descriptive framework with categories of technical and non-functional impacts were submitted in response to this puzzle. The results show that the application of the SIE, allows having a founded notion of the level and scope of the SaaS impacts in the organization, allowing the decision maker to select solutions involving a minor impact and/or develop strategies to mitigate the impacts identified. The SIE and the descriptive framework try to instill formality, objectivity and rationality on the SaaS evaluation process, requiring the assessor to ask questions directed at each impact category, looking simultaneously to the impacts not only in their organization, as a consumer, but also in the SaaS provider organization, or vice versa. The SIE can be considered a normative decision model that allows the business strategy decision maker to make decisions to maximize value for their organization.

The application of SIE to this study identified what should be the focus of attention of software developers, service providers, users and also researchers towards redefining business and IT strategies, and also reevaluate investigation lines. To implement the first step, we present the following tool that can help to identify the SaaS impacts by filling in a table where each column represents different groups of stakeholders and each line represents different categories of impact. Each cell represents the impact of a specific group in a specific category. A company that wants to identify its relevant impacts should fill one of the cells of this table identifying impacts on their organization. Bellow we presented the summary of the main impact categories that were identified in the present study and that influence the business model, the application architecture and operational structure and reflect the constructed descriptive framework of SaaS impacts:

**Quality of service:** Fit for purpose (functionality) and suitable for use (availability and warranty). To measure the quality level of SaaS, providers generally are based on the levels of security, availability, compliance, service levels and trust that customers place in service.

**Benefits:** Key benefits that are available in Software as a Service.

**Reliability, market positioning and SLAs:** Related to the credibility that the company that is providing the service has on the market - what references it has in the market. Terms and conditions of SLAs negotiated between the parties - levels for service availability, performance, support, etc., and penalties in case of the agreement failure.

**Revenue and Profit:** As we see in this study, the cost reduction is one of the main reasons for SaaS adoption. The return of the investment in SaaS and perspectives for the future, may be a decision factor in certain cases.

**Payment Model:** Payment options or lack thereof can be effectively a difficulty in the design and development of SaaS and also a limitation to the customer. Among the options we can speak of models based on subscription (yearly, monthly, etc.), based on the use (e.g. based on number of transactions, sales, etc.).

**Time to Market:** The time-to-market is very important for all businesses. The SaaS offering is not an option for anyone who delivers software today and who wants to keep their position on the market has to follow the trend and give this option to their customers.

**Implications:** The movement for this new paradigm may have several implications and difficulties that companies have to overcome: May require internal reorganizations, strategies changes, skills updates, etc.

**Functionality levels, updates and roadmap:** Providing different functionality levels, upgraded software through various monthly, quarterly, or annual updates as new releases are ready. Provide services in a roadmap perspective of evolution and service continuity.

**Customization:** Allows the customer to customize the application according to their specific needs without interfering with the core of the application, not impacting the processes of the other customers and the correct operation of the application.

**Integration:** Providing integration interfaces and processes that allow the customer to make the integration of SaaS application to their internal systems. The solutions enable interoperability and integration with other applications in the organization and it may be an important factor in the evaluation and selection of the software service.

**Architecture:** Single-tenant, multi-tenant application and/or database. The availability of the service in Online and Offline hybrid models may be a need in the event that connectivity and access is a problem.

**Security:** Protocols, communication and data security tools. The need for encryption, ensuring information security and integrity has implications on the solution architecture.

**Availability and performance:** Providing of disaster recovery processes and replication. Elasticity of the infrastructure and also the application itself, to answer adequately to the customer´s needs.

**Virtualization and infrastructure model:** Providing services in virtualized environments or not. What infrastructure models are available, in-house, External, PaaS, IaaS can identify new implications in dependence on third parties and the associated sub-contracted service quality conditions.

**Technologies:** Technologies and development environments used may have operational and technical implications.

As an example, we present below some cells filled with impact statements on a specific group in a specific category:

| | | Customers | Service providers | Developers |
|---|---|---|---|---|
| **Non-functional requirements** | **Quality of service** | Business Processes | Opportunity growth higher costs | Robustness of the application |
| | **Benefits and efficiency** | Colaboration | Productivity | Time-to-market |
| | **Reliability** | Security Business Processes | Acquiring new customers | Support, fixes Releases |
| | **Cost-cutting** | Subscription /Use Fewer licenses Resources Least h/w | Opportunity Economy of scale | Use Complex model |
| | **Payment model** | Setup cost and periodic productive economy | Monitoring and control tools | Applicational complexity Functions |
| | **Time to market** | Market positioning New business | Market performance Competitive edge | Innovation Competitive edge |
| | **Implications** | HR resources Functions Reallocation Processes changes | Resources New skills | Resources New skills |
| | **Satisfaction** | Choice of supplier | Competitiveness Image | Competitiveness Image |
| **Technical requirements** | **Functionality levels, updates and roadmap** | Costs according to functionality level | Control Tools | Control Tools |
| | **Customization** | Business requirements | Architecture Support and maintenance | Architecture Technologies |
| | **Integration with other applications** | Productivity | Availability Security | Interfaces |
| | **Architecture** | Security and information privacy | Opportunity to lower costs | DB Multi-tenant App Multi-tenant |
| | **Security** | Confidence Business processes | Agreements with third parties | Encryption Protocols |
| | **Availability** | Revenue and Profit | Infrastructure Redundancy, DR | BD Archtecture Sessions |
| | **Virtualization** | Availability Quickly to restore service | Cost-cutting | |
| | **Tecnologies** | Integration with internal systems | New skills | New skills |
| | **Trend** | Social | Big data | Mobility |

**Tab. 15 – Impact identification matrix**

All actors in this theater of services, whether they are manufacturers Software, suppliers or customers, want to create value with the service. The customer wants to take advantage of this value without having responsibility for specific costs and risks associated with providing the service that is held by the service provider [48]. Thus, based on the fundamental concepts of PMBOK risk management and ITIL is essential to identify and assess the impact that the new service will have on business, in order to mitigate or eliminate the limitations and enhancing the benefit. Risks can be positive or negative. As a chemical reaction, we can identify the correct catalyst to mitigate or eliminate the negative risk, which was a challenge, and also to enhance or accelerate the positive risk as an opportunity. After the identification of impacts it is necessary to classify them. According to the PMBOK, each risk is evaluated according to the probability and the impact of the risk [47]. The organization shall determine the combinations of probability and impact in order to classify high risk, moderate risk, and low risk, but the classification rules differ and are specific to each organization or project. Based on this concept, we propose the development of a matrix according to the probability and magnitude:

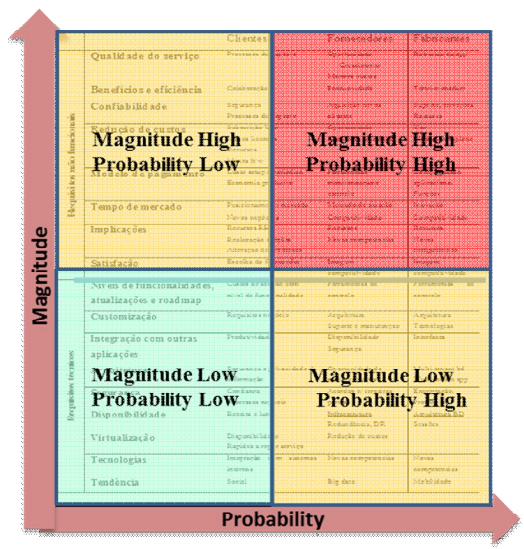

Fig. 301 - Impact Classification matrix

In step 2 of our proposed process the survey is planned and designed. Let's consider, for example, we are a SaaS developer and provider, and the customization option for the customers had a high impact on my SaaS offer. What kind of issues could I put to us as a service provider and also to our customers? Because the customization option for each client could have implications on the architecture of the software and also of the hardware solution, the following questions could be included:

| Category | Domain | Questions |
|---|---|---|
| Customization | Developer/Provider | 1. List the factors that you think are the main categories of concern in the development of SaaS |
| | | 2. Do you manage the infrastructure of hardware / software (servers, storage, security, software, etc.) for your SaaS application? |
| | | 3. If you answered "Outsource" in the previous question, what is the main type of outsourcing that you use? |
| | | 4. What percentage of IT budget you spent on maintaining this infrastructure? |
| | | 5. Rate each option of developing and delivering SaaS based on real cost savings in the future? |
| | | 6. Do you give an API (application programming interface) documented for your application to your customers and/or third parties? |
| | | 7. Do you use third party components such as Boomi, Cast Iron, Pervasive, etc. to automate data integration and business processes in your SaaS applications? |
| | | 8. Which of the following statements best describes the architecture of your SaaS software? |
| | | 9. If Multi-tenancy is not answered in the previous question, what was/is the main reason not to implement an architecture multi-tenancy in your SaaS system? |
| | | 10. Are you developing a SaaS product on a platform (PaaS) from third parties such as Force (Salesforce.com), NetSuite, or other similar system? |
| | Customers/Users | 1. Indicate the advantages of using SaaS solutions rather than on-premise software (installed locally) |
| | | 2. What factors do you think are the main areas of concern in the use of SaaS? |
| | | 3. Do you have customized SaaS applications? |
| | | 4. How do you compare the experience of customization of SaaS applications compared to traditional software? |
| | | 5. After customization, how do you compare the capabilities and functionality of the SaaS application with the traditional software? |
| | | 6. Which SaaS Providers are you using? |

Step 3 identifies the inquired sample that should involve several people from different areas of the organization who provides the service, to ensure that the information is reliable and consistent, in response to the questionnaire for Developer/Provider. To answer the questionnaire for Customers/Users, several different people in the market that we intend to address should be selected.

Step 4 runs the questionnaires.

Step 5 proceeds to the collection of information. This should be stored in a safe and central and accessible location for further analysis by the various stakeholders involved in the interpretation.

Step 6 proceeds to the generation of the report in which the results with the answers to the questions putted to the various stakeholders, making clear the impartiality of the evaluator.

Step 7 proceeds to the interpretation of the results, indicating the positive and negative impacts of the recommendation and presents a recommendation to facilitate the decision making.

## 5. CONCLUSION AND CONTRIBUTIONS

By applying the SaaS evaluation framework to the present study, we took the following main conclusions:

*SaaS customers should set their own strategies for integrating SaaS and should incorporate them in a holistic approach to multi-company integration in all projects [34].* This study confirms this Lheureux assertion. However, there are many limitations for integration into analyzed SaaS offering.

*The integration and customization are critical components in the strategies of successful SaaS architecture, in an IT infrastructure centralized on services [7].* This study confirms this statement. However, there are many limitations in customization of analyzed SaaS offering.

The results of this study imply that managers of Portuguese companies recognize the added value and surplus value of the Software-as-a-Service in their organizations. However, there are some needs for modification, adaptation and evolution. SaaS is not only a different software distribution model, it is a new way of doing business with the software. Nowadays, companies have at their fingertips the technology that allows them to embrace new markets and new business and should identify business strategies, organizational changes and needs of human resources such as acquiring new skills and/or reorganization of functions.

The added value of this study lies mainly in the integration of the two perspectives in terms of who provides and who consumes the software to evaluate the impact of a SaaS transition and provide an understanding of the real needs of the Portuguese companies that can be generalized to other geographies. Moreover, the proposed framework to guide SaaS developers, providers and customers, is a useful resource to assess the impact of SaaS in their organizations. A proposal of tools that help the effectively assess of specifications to meet real business needs, allow to take more assertive decisions based on sustained information. We hope that the research results can provide information for decision-making on the SaaS adoption within strategic lines of IT organizations and can be useful in the discussion or clarification of technological and organizational issues needed to support the data, information and business processes.

This research has opened new clues for future studies. Among the possible research lines that emerge from this study, for their relevance and interest, are the following:

- Applying the proposed SIE framework, in the evaluation of a new SaaS service implementation in the organization, streamlining the decision process by reducing uncertainty and facilitating the decision making;
- Complement the proposed framework with strategy business processes, solution design and decision making models;
- Research on the impact of SaaS framing different perspectives turned out to be a relevant approach with contributions from different markets. However, we restrict our analysis to Portugal. Future research should replicate this research project to other countries to get an overall picture about the impact of SaaS in the global market.